# Using Multimedia Presentations to Improve Digital Forensic Understanding: A Pilot Study

**Niken Dwi Wahyu Cahyani**
School of Information Technology and Mathematical Sciences
University of South Australia
Adelaide, Australia
Email: niken.cahyani@mymail.unisa.edu.au

**Ben Martini**
School of Information Technology and Mathematical Sciences
University of South Australia
Adelaide, Australia
Email: ben.martini@unisa.edu.au

**Kim-Kwang Raymond Choo**
School of Information Technology and Mathematical Sciences
University of South Australia
Adelaide, Australia
Email: raymond.choo@unisa.edu.au

## Abstract

Improving employees' understanding of digital forensic technical terms and concepts within an organisation is likely to increase the potential of successful collaboration during a cyber security incident (e.g. data breach) investigation within that organisation. In this paper, we seek to determine whether multimedia presentations, in this case videos, are an effective tool in improving a learner's technical understanding of digital forensic terms and concepts. Using the cognitive theory of multimedia learning as the underlying theoretical lens, we surveyed nine participants from the financial sector who have cyber security-related responsibilities. With the exception of one participant, the study found that the use of multimedia presentations can improve participants' understanding of technical digital forensic terms and concepts. Potential future research questions are also identified.

**Keywords**

Comprehension, multimedia presentations, digital forensic training, multimedia learning, financial services.

## 1 Introduction

Cyberattacks and cybercrime, including those against financial services institutions, are becoming more frequent, sophisticated and cross-border (Choo 2011; Cuomo and Lawsky 2014). In the investigation of a cybersecurity incident, identification, preservation, analysis and presentation of evidence (a process known as digital forensics – McKemmish 1999) can be used to guide an incident response investigation that seeks to answer six key questions of the incident (i.e. what, why, how, who, when and where).

For this reason, an understanding of digital forensics is essential to both law enforcement and private sector organisations (Ab Rahman and Choo 2015a; Nance et al. 2010; Quick et al. 2013; Yasinsac et al. 2003). For the latter, incident response and investigation processes can be expedited if organisations have a sound understanding of incident response and digital forensic principles (Ab Rahman and Choo 2015b; Nance et al. 2010), and their incident response team is forensically trained. This need has been covered in the academic literature, with scholars, such as Ab Rahman and Choo (2015a), Kessler (2011) and Vidal and Choo (2015), highlighting the need for incident responders and investigators to have an improved understanding of digital forensics. Education and training programs are, therefore, necessary to equip incident responders and investigators with the skills to assess both the digital evidence presented and the digital forensic investigation that was conducted, specific to the case.

Multimedia presentations, commonly characterised by narration and animation (Mayer and Moreno 2002), include a variety of distinct media such as text, image, audio and video (Prabhakaran 2000). Mayer emphasised the role of multimedia presentations as an effective training method, as the combination of spoken text and animation can result in a deeper understanding for learners (Mayer 2003). Multimedia presentations have been identified as effective in empowering learners' understanding of business materials in sectors, such as banking and finance (Gunasekaran and Love 1999), as well as explaining concepts such as DNA (Hewson and Goodman-Delahunty 2008; Schmitt 2007) and restorative justice (Varfi et al. 2014). In an earlier work, we sought to determine the utility or effectiveness of multimedia presentations in digital forensics training. We conducted a questionnaire-based survey using a convenient sample of judges, investigators, prosecutors and staff from three provinces in Indonesia. Our findings showed that all participants had an increased level of understanding after viewing the educational videos (Cahyani et al. 2016).

In this paper, we extend our previous study to include participants from a different industry sector, namely: the financial sector, seeking to answer the following question: "Are multimedia presentations (i.e. videos) an effective tool to improve a learner's technical understanding of digital forensic terms and concepts?".

## 2   Cognitive Theory of Multimedia Learning

The cognitive theory of multimedia learning (Mayer and Moreno 2002) identifies two channels in the human information processing system, representing verbal (narration) and visual (animation) channels. According to the theory, learners will identify aspects of the words in the narration via the verbal channel, and other aspects will be pictures in the animation through the visual channel. Then, learners will organise these words and pictures so that they can be correlated. Finally, learners will build a connection with their prior knowledge in order to learn new knowledge based on what they have just received. The main concept of the cognitive theory of multimedia learning is depicted in Figure 1.

This theory is useful as the underlying theoretical lens in designing multimedia presentations for laypersons (i.e. untrained or unfamiliar with digital forensic terminologies). As depicted in Figure 1, the content materials are scientific text and animation.

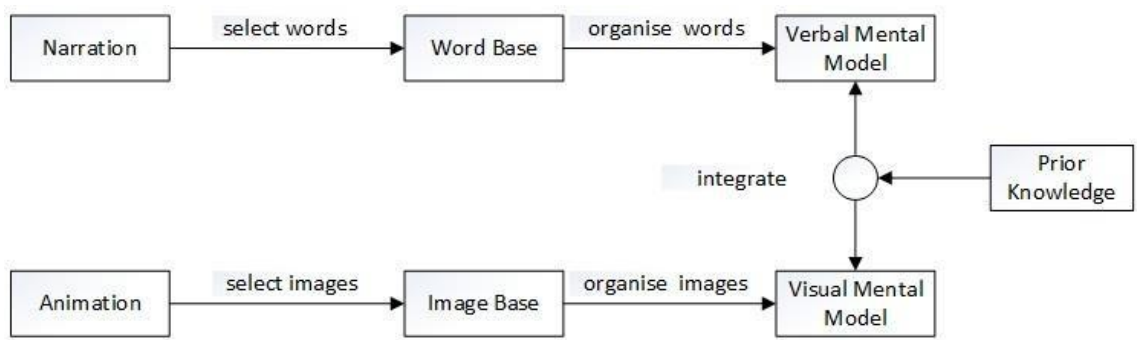

*Figure 1: A Cognitive theory of multimedia learning (adapted from Mayer and Moreno 2002, p. 111)*

In the theory, Mayer and Moreno (2002) recommend five aids to computer-based multimedia learning (see Figure 2). These aids represent principles in instructional design to maximise the benefits of multimedia presentations. For this study, the multimedia and contiguity principles were adopted in the multimedia presentations because they were found to be the two most significant principles in the improvement of a learner's understanding by Mayer and Moreno (2002). Meanwhile, the coherence, modality, and redundancy principles were adapted, which allow us to emphasise on the use of sounds and on-screen text.

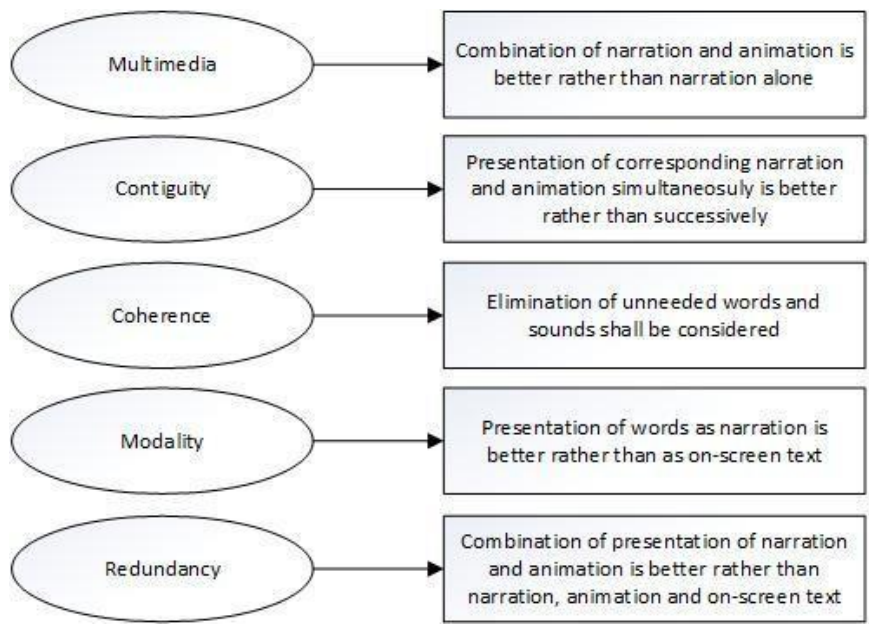

*Figure 2: Five aids to computer-based multimedia learning (adapted from Mayer and Moreno 2002)*

## 3 Research Methodology

### 3.1 Participants

The sample consisted of nine participants who attended the "Cyber Resilience in Financial Institutions" workshop conducted by the last author in Singapore during March 2015. These participants held senior positions with cyber security-related responsibilities, and their socio-demographic information is listed in Table 1.

Three of the participants (33.3%) were aged between 26 and 35 years, and the remaining participants were aged between 36 and 55 years. Eight of the participants (88.9%) were from East Asia (Malaysia, Myanmar, Singapore, Vietnam) and one participant was from Spain. Only four participants were native English speakers, and only one participant had not obtained at least a Bachelor's degree. All participants work in the financial sector. The participants were also asked to self-rate their technical literacy level (1 is the lowest and 5 is the highest).

| Participant | Age Range | First Language | Highest Education | Job Title | Technical Literacy |
|---|---|---|---|---|---|
| 1 | 36-45 | Non-English | Master's degree | Director | 4 |
| 2 | 26-35 | English | Bachelor's degree | Manager | 4 |
| 3 | 36-45 | English | Bachelor's degree | Vice-President | 4 |
| 4 | 36-45 | English | Master's degree | Director | 3 |
| 5 | 46-55 | Non-English | Diploma or lower | Manager | 2 |
| 6 | 26-35 | Non-English | Bachelor's degree | Senior Manager | 3 |
| 7 | 36-45 | Non-English | Bachelor's degree | Deputy Managing Director | 4 |

| 8 | 26-35 | Non- English | Master's degree | Manager | 4 |
| 9 | 46-55 | English | Master's degree | Director | 5 |

*Table 1. Participants' Background Information*

### 3.2　Interview Materials

The questionnaire, adapted from our previous study (Cahyani et al. 2016), was divided into three parts. In the first part, we surveyed the participants' socio-demographic background. In the second part, we asked the participants to provide a definition for the three technical terms common to cyber security or digital forensic investigations, namely "Cloud Computing", "Botnet", and "Forensic File Recovery". The participants were advised to answer "unfamiliar" if they were not familiar with the term.

The term cloud computing was chosen due to the increasing adoption of cloud services by financial institutions (Nicoletti 2013). As such, an understanding of cloud technical concepts is critical for both investigators and incident responders (Choo 2010; Rani and Gangal 2012). Botnets are commonly used in cyberattacks, including against cloud service providers. For example, specific botnets, also known as financial botnets, are built to carry out financial fraud (Riccardi et al. 2010). Therefore, botnet was one of the three terms used in our study. The last term, forensic file recovery, was selected as it is a common yet important concept in many digital forensic investigations.

After the participants had provided their own definitions of the terms, they were then shown the multimedia presentations, one for each term. After viewing the presentations they were requested to once again define the terms.

The final part of the interview consisted of two open-ended questions, where participants were able to rate and comment on the usefulness (or limitations) of the multimedia presentations in enhancing their understanding of the three technical terminologies and concepts, and the multimedia presentation format and content. They also had the opportunity to recommend any other tool(s) or method(s) that they think would facilitate their understanding, and provide any further comments regarding the study.

The multimedia presentations were originally developed at our University, and were chosen or customised to fit the five aids proposed by Mayer and Moreno (2002).

Aid 1. Multimedia –The first and second videos are narrated by a male voice, and the third video is narrated by a female voice, to provide voice and gender variety. Animations are used in the videos. Video 1 first introduces the cloud definition, followed by a selection of implementations. Similarly, in video 2, the term botnet is first defined, followed by an explanation of how they are used in a cyberattack or cybercrime. In the last video, the sequence of events that occurs in a file deletion and recovery process is explained.

Aid 2. Contiguity –In the selected videos, when an animation is shown, a supporting narration is also played simultaneously. However, the video delivery was designed so that when it is necessary to emphasise the explanation given in the narration, the animation can be paused.

For example, in video 1 when a narration explains what a client can access via a cloud service, the animation shows the same frame until the explanation is completed.

Aid 3. Coherence –Since the videos' durations are relatively short (around three to five minutes), narration and word selection was especially considered in order to provide proper information to support the relevant animation. For example, in video 2, words are selectively chosen to explain the complex botnet formation process and its attacks. However, in all three videos, appropriate background music is used to maintain the attention of the learner.

Aid 4. Modality –This principle in generally adopted in all three videos. However, we particularly selected videos that use on-screen text in a limited way. We are aware of the importance of integrating multiple sources of information (e.g. separate text and diagrams) to reduce cognitive load (Chandler and Sweller 1991). As such, on-screen text is utilised to emphasise the explanation only when needed. For example in video 3, when the file deletion process is explained narratively, on-screen text about the process that has been applied to the file is displayed to assist the learner in comprehending the sequence of activities conducted and their impact, see Figure 3.

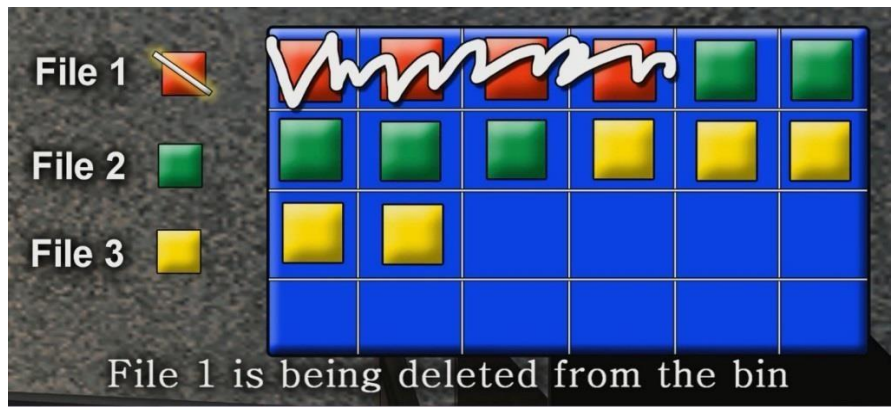

*Figure 3: An example of on-screen text usage (video designed by Kwan Wong and Robb Weston, University of South Australia)*

Aid 5. Redundancy – Presentation of narration, animation and on-screen text is used simultaneously in selected frames to remind the learner about a particular process stage.

### 3.3 Procedure

We used an offline questionnaire, which was manually distributed to the workshop's participants. One hour was allocated for the study. Upon completion, the participants' responses were then manually entered, coded and analysed by the researchers.

In order to grade the participants' responses to the three terms before and after watching the multimedia presentations, we used a grading metric. For example, in the cloud computing video, cloud computing is defined in terms of the deployment models, the associated benefits, and the architecture/service models. If participants only defined cloud computing as a shared pool of configurable computing resources/virtualized computing resources, then we would award them a maximum of 0.5 marks. Zero marks were assigned only to those who answered unfamiliar or did not answer the question. The maximum score for each answer was 3.

We also computed an ability to comprehend value ($\bar{x}$) for each participant to determine the extent of the impact of the multimedia presentations on their understanding. Similar to our earlier work (Cahyani et al. 2016), the participant's ability to comprehend was calculated using the following formula:

$$\bar{x} = \frac{(x_{v1\_Af} - x_{v1\_Bf}) + (x_{v2\_Af} - x_{v2\_Bf}) + (x_{v3\_Af} - x_{v3\_Bf})}{3}$$

In the above formula, $x_{v1\_Af}$, $x_{v2\_Af}$ and $x_{v3\_Af}$ denote the 'participant's understanding' after watching videos 1, 2 and 3 respectively while $x_{v1\_Bf}$, $x_{v2\_Bf}$ and $x_{v3\_Bf}$ denote the 'participant's understanding' before watching videos 1, 2 and 3.

## 4 Results and Analysis

This section presents the findings of the pilot study.

### 4.1 Multimedia Presentation Impact on Participants' Understanding

Our grading of the participants' responses before and after watching the multimedia presentations is listed in Table 2. It appears that the participants' understanding of term 3 was the lowest before they watched the videos but significantly improved after watching video 3. With the exception of Participant 1, the participants' understanding of the terms improved after viewing the presentations.

| Participant | Before watching the video | | | After watching the video | | |
|---|---|---|---|---|---|---|
| | Term 1 | Term 2 | Term 3 | Term 1 | Term 2 | Term 3 |
| 1 | 2.13 | 1.63 | 1 | 2.13 | 1.63 | 1 |
| 2 | 0.38 | 0.38 | 0.75 | 1.25 | 1.58 | 2.38 |
| 3 | 0.50 | 1.50 | 1.25 | 0.88 | 2 | 1.88 |
| 4 | 0.75 | 1 | 0.50 | 2.38 | 2.13 | 1.75 |
| 5 | 0 | 0 | 0 | 0.50 | 0.50 | 0.50 |
| 6 | 1.43 | 1.63 | 0.25 | 2.38 | 2.13 | 2 |
| 7 | 1.13 | 0.75 | 0.63 | 2.38 | 1.88 | 2.63 |
| 8 | 0.63 | 0.75 | 0.75 | 1.75 | 1.13 | 2.25 |
| 9 | 0.88 | 1 | 0.75 | 1.25 | 2 | 1 |

*Table 2. Participants' understanding of the three terms*

We also computed the mean of the differences between the 'After watching the video' and 'Before watching the video' comprehension values for each of the terms (i.e. the average of the 'After watching the video' value minus the 'Before watching the video' value). In this study, we found that participants' comprehension of the three terms before watching the videos was 0.83 with a standard deviation of 0.44 and after watching the videos was 1.67 with a standard deviation of 0.53. The mean difference value before and after watching videos is 0.85, with a standard deviation of 0.48.

Each participants' ability to comprehend is depicted in Figure 4. A positive value (>0) for ability to comprehend indicates an improved understanding after watching the videos.

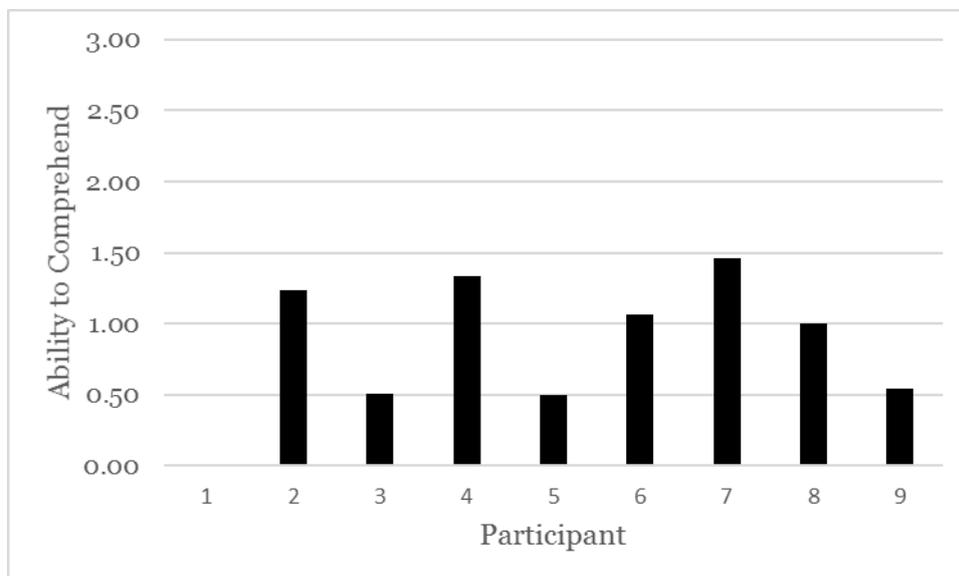

*Figure 4: Participants' ability to comprehend*

## 4.2 Participants' Feedback

Participants were also asked to rate the usefulness of the multimedia presentations in enhancing their understanding of the technical terms and concepts. The rating is presented in Figure 5. Eight participants (88.9%) gave a rating of 3 or more, and three of the eight participants gave the highest rating of 5.

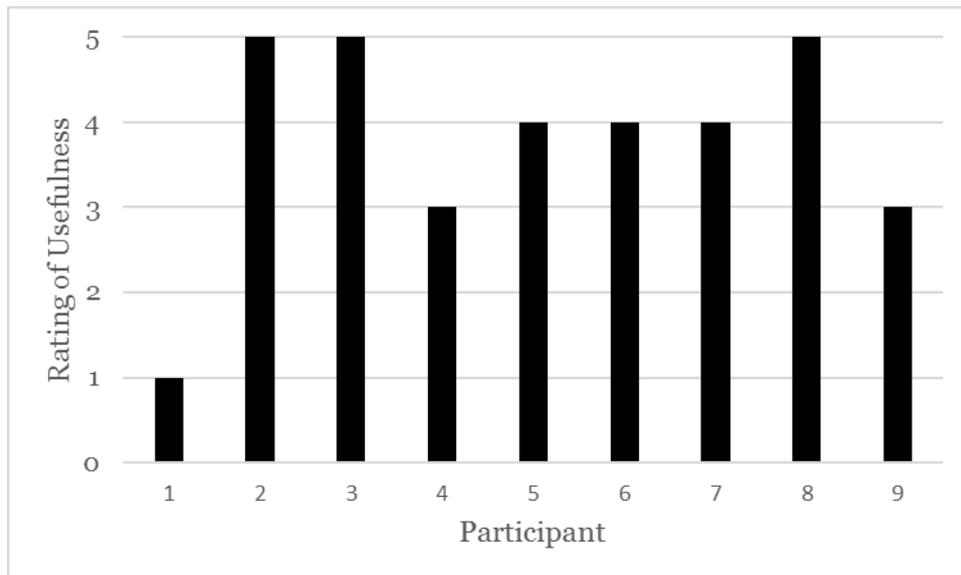

*Figure 5: Rating of multimedia presentation usefulness*

The participants' feedback is listed in Table 3. Participant 1, who experienced no improvement in their understanding of the terms and concepts after watching the videos, gave a rating of 1 (the lowest). They went on to explain that the three terms selected for the study were 'basic concepts'. Other participants (e.g. Participants 5 and 6), however, stated that the videos were useful in enhancing their understanding of the materials.

| Participant | Feedback |
|---|---|
| 1 | Those [terms are] … basic concepts. |
| 2 | [I recommend e]xplanation using analogy in everyday life [and] hands-on demonstration. |
| 3 | [I recommend v]ideo conference/talk (recording) [and] share current practices by various industries/countries. |
| 4 | …using video … [and] pointing to website with actual presentation. |
| 5 | [The videos are s]imple and easy to understand. [I recommend s]imilar approach for Team Building Events with activities and role play. |
| 6 | I think video is the effective tool. |
| 7 | [I recommend d]emonstration of processes based on technical terminologies. |
| 8 | [Recommendation is d]ependent [on the] situation: for network [terms and concepts, I recommend] lab and diagrams for forensic [terms and concepts, I recommend] lab and bullet points for software [terms and concepts, I recommend] Powerpoints and bullet points |
| 9 | [no feedback was provided] |

*Table 3. Excerpts from the participants' feedback*

## 5　Discussion

It is no longer a question of if malicious cyber activities will negatively affect a business, but of when, and consequently, the role of digital forensics is now more important than ever. It is, however, challenging for technicians and practitioners to keep pace with technological advancements (and the associated terminologies and concepts). Therefore, it is important to devise an effective means for improving an individual's understanding of complex digital forensic terms and concepts.

While the findings in this study, as well as our previous study (Cahyani et al. 2016), on the ability of multimedia presentations to improve an individual's understanding of complex digital forensic terms and concepts have been promising, both pilot studies were limited by their sample size. Therefore, it is essential to extend this research to include participants from different sectors on a wider scale. In addition, the four components identified by Taylor et al. (2007) may be useful in both the choice of terms and design of future multimedia presentations:

Component 1: Multi-disciplinary content and, therefore, a wider range of contents to be covered. In the future, we will canvass the opinions of key stakeholders in digital forensics and incident response, particularly those in law enforcement, academia, the private sector, and the judiciary, to obtain a list of key technical terms that are common in incident response and forensic investigations, judicial or administrative proceedings.

Component 2: Hands-on exercises. Participants had also commented that hands-on exercises might be a more effective way of learning. Therefore, in future research, we will also examine the merits of other education platforms. For example, due to the increasingly popularity of mobile devices, would it be possible to leverage mobile learning platforms (Lu and Viehland 2008)?

Component 3: Knowledgeable instructors. In our pilot studies, we did not distinguish between beginner, intermediate and advanced participants. In future research, we will examine the need to design materials suited for different participant groups.

Component 4: Real world case examples. As the terms and concepts used in digital forensics might be new for some participants, the use of analogy may help to facilitate participants' understanding, particularly for participants with little or no technical knowledge.

## 6　Concluding Remarks

In this pilot study, we examined the potential for using multimedia presentations to enhance participants' understanding of technical terms and concepts in digital forensics. We found that eight out of the nine participants (88.9%) in this study experienced an increase in their level of understanding after watching the multimedia presentations. This suggests that multimedia presentations are likely to be an effective tool in improving a learner's understanding of digital forensic terminologies. The findings in this study echoed our previous observations in a study involving participants comprising judges, investigators, prosecutors and staff from three provinces in Indonesia (Cahyani et al. 2016).

## 7　References